# An experimental and theoretical investigation of the C($^1$D) + D$_2$ reaction

*Kevin M. Hickson*[*,†,‡] *and Yury V. Suleimanov*[*,∥,^]

[†] Université de Bordeaux, Institut des Sciences Moléculaires, F-33400 Talence, France

[‡] CNRS, Institut des Sciences Moléculaires, F-33400 Talence, France

[∥] Computation-based Science and Technology Research Center, Cyprus Institute, 20 Kavafi Str., Nicosia 2121, Cyprus

[^] Department of Chemical Engineering, Massachusetts Institute of Technology, 77 Massachusetts Ave., Cambridge, Massachusetts 02139, United States

**AUTHOR INFORMATION**

**Corresponding Authors**

km.hickson@ism.u-bordeaux1.fr, ysuleyma@mit.edu

**Abstract**

In a previous joint experimental and theoretical study of the barrierless chemical reaction C($^1$D) + H$_2$ at low temperatures (300-50 K) [K. M. Hickson, J.-C. Loison, H. Guo, Y. V. Suleimanov, *J. Phys. Chem. Lett.*, 2015, **6**, 4194.], excellent agreement was found between experimental thermal rate constants and theoretical estimates based on ring polymer molecular dynamics (RPMD) over the two lowest singlet potential energy surfaces (PESs). Here, we extend this work to one of its deuterated counterparts, C($^1$D) + D$_2$, over the same temperature range. Experimental and RPMD results are in very good agreement when contributions from both PESs to this chemical reaction are included in the RPMD simulations. The deviation between experiment and the RPMD calculations does not exceed 25 % and both results exhibit a slight negative temperature dependence. The first excited $^1A''$ PES plays a more important role than the ground $^1A'$ PES as the temperature is decreased, similar to our previous studies of the C($^1$D) + H$_2$ reaction but with a more pronounced effect. The small differences in temperature dependence between the earlier and present experimental studies of C($^1$D) + H$_2$/D$_2$ reactions are discussed in terms of the use of non-equilibrium populations of ortho/para-H$_2$/D$_2$. We argue that RPMD provides a very convenient and reliable tool to study low-temperature chemical reactions.

**1 Introduction**

Gas-phase reactions involving neutral carbon atoms are considered to be important in combustion systems, in the chemistry of planetary atmospheres and in the interstellar medium. Many experimental and theoretical studies [1-13] of the reactivity of ground state C($^3$P) atoms have been performed, addressing their kinetic and dynamic aspects over a wide range of temperatures [2-6, 8, 9] and collision energies [9-13] and for a variety of different coreagent molecules. The reactivity of the excited $^1$D state of atomic carbon is generally less well understood, with the exception of the C($^1$D) + H$_2$ reaction; a process which is simple enough to allow comparison between precise state-selected experimental measurements [14-16], quasi classical trajectory calculations [17-19] and detailed quantum mechanical methods, [20-29] even for several isotopic forms of the H$_2$ molecule (H$_2$, HD, D$_2$) [18, 19, 30-32]. Although there are several experimental works following the dynamics of other C($^1$D) atom reactions under a range of conditions [13, 33-36], few kinetic investigations are reported in the scientific literature. Of these, the majority is confined to room temperature [37-42]. Indeed, the only temperature dependent measurements of the rate constants for C($^1$D) reactions to have been performed to date are those between C($^1$D) + CH$_3$OH [7], C($^1$D) + H$_2$ [43] and the non-reactive quenching of C($^1$D) atoms with N$_2$ [44]. Extending such studies to include other isotopic forms of the same molecule would allow us to look for interesting reactivity differences, thereby providing a rigorous test of the underlying potential energy surfaces (PESs).

Previous room temperature kinetic studies of the C($^1$D) + H$_2$, HD and D$_2$ reactions [42] demonstrated a noticeable isotope dependence of the rates of these processes, with the rate constant decreasing from H$_2$ to HD to D$_2$. Lin and Guo [24] were able to reproduce the experimental results of Sato *et al.* [42] using a wavepacket based statistical model over the lowest $^1A'$ PES, attributing the reactivity difference to a kinematic effect related to the increasing reduced mass. The recent joint experimental and theoretical investigation of the C($^1$D) + H$_2$ reaction by Hickson *et al.*[43] has cast some doubt on the reliability of these earlier experiments, measuring rate constants which are significantly larger than those measured by Sato *et al.*[42]. On the theoretical side, novel rate theory based on ring polymer molecular dynamics (RPMD)[45] was implemented that demonstrated very reliable and predictable behavior in multifarious prototype cases as discussed in a recent review[46]. Using the RPMD method, these authors[43] were also able to demonstrate that the new experimental results could be well described if reaction occurred adiabatically over both the lowest $^1A'$ and first excited $^1A''$ PESs, considering these surfaces to be uncoupled. Interestingly, Defazio *et al.* also studied the quantum dynamics of the C($^1$D) + H$_2$, HD and D$_2$ reactions over uncoupled $^1A'$ and $^1A''$ PESs using a quantum mechanical time dependent wavepacket method [22, 32]. They found that while the upper $^1A''$ plays an important role in all these reactions, the reactive contribution of this surface is particularly large in the case of C($^1$D) + D$_2$. Clearly, an investigation of the temperature dependent kinetics of the reactions of C($^1$D) atoms with other deuterated forms of hydrogen would provide an ideal opportunity to validate the earlier hypotheses of both Defazio *et al.*[32] and Hickson *et al.*[43] with regard to the role of the $^1A''$ PES in these systems.

Here, we report the results of an experimental and theoretical investigation of the C($^1$D) + D$_2$ reaction at low temperature. A supersonic flow reactor employing pulsed laser photolysis for C($^1$D) generation coupled with vacuum ultraviolet laser induced fluorescence detection of the deuterium atom products allows us to follow the reaction kinetics over the 50-296 K range. In parallel, RPMD simulations were carried out for the title reaction in the same temperature range. Sections 2 and 3 describe respectively the experimental and theoretical methods used in this work and the results which are further summarized in section 4.

## 2 Experimental Results

Measurements were performed using a continuous flow reactor, employing axisymmetric Laval nozzles to generate the cold supersonic flows. As this method is well described elsewhere [47, 48], only features specific to the current work will be presented here. For the present investigation, it was impossible to use any of our Laval nozzles based on $N_2$ as the carrier gas, as the $C(^1D) + N_2$ quenching reaction has been shown to become more efficient as the temperature falls [44]. As a result, three nozzles employing the carrier gas argon were used, allowing temperatures of 50 K, 75 K and 127 K to be attained. Details of the flow characteristics of these three nozzles can be found in Table 1 of Grondin *et al.* [49]. By removing the Laval nozzle and by reducing the flow velocity, it was also possible to perform room temperature measurements (296 K) with argon as the carrier gas.

$CBr_4$ was used as the source of $C(^1D)$ atoms in these experiments. These precursor molecules were carried into the reactor by a small argon flow passing over solid $CBr_4$ held in a separate vessel at a known fixed pressure and temperature. This allowed us to estimate the concentration of $CBr_4$ molecules within the supersonic flow to be less than $2 \times 10^{13}$ molecule cm$^{-3}$. The pulsed multiphoton dissociation of $CBr_4$ at 266 nm produced both $C(^1D)$ and $C(^3P)$ atoms along the entire length of the supersonic flow [4-8, 44]. Although the major photolysis product is $C(^3P)$ atoms [7], the $C(^3P) + D_2 \rightarrow CD + D$ reaction can be neglected at room temperature and below due to its high endothermicity of 120 kJ mol$^{-1}$. Due to inefficient gas-phase spin conversion, the $D_2$ used in the present experiments was characterized by a fixed ortho/para ratio of 2:1 at all temperatures. This can be contrasted with the expected value at 50 K, which should approach 4:1 assuming thermal equilibrium is attained. The $D_2$ concentration was held in large excess with respect to $C(^1D)$ atoms in this work, so that pseudo-first-order conditions could be assumed. As $C(^1D)$ atoms could not be followed directly in the present experiments, product D($^2$S) atoms were detected instead through vacuum ultraviolet laser induced fluorescence (VUV LIF) at 121.534 nm. Light at this wavelength was generated by frequency doubling the 729.2 nm output of a pulsed dye laser, yielding a UV beam at 364.6 nm, which was then focused into a cell containing 210 Torr of Kr and 540 Torr of Ar, producing the required tunable VUV light by third harmonic generation. A magnesium fluoride lens positioned at the exit of the tripling cell allowed the VUV beam to be collimated and steered into the reactor where it was allowed to cross the supersonic flow. The on-resonance VUV emission from excited D-atoms within the flow were collected by a lithium fluoride lens and focused onto the photocathode of a solar blind photomultiplier tube.

None of the gases used in the experiments (Ar 99.999%, D$_2$ 99.8%, Kr 99.99%) were purified prior to use. Instead, these gases were flowed directly from cylinders into mass-flow controllers which provided precise control over the carrier gas, reagent gas and precursor concentrations within the supersonic flow. All mass-flow controllers were calibrated using the pressure rise at constant volume method for each individual gas used.

Typical product D($^2$S) VUV LIF intensity profiles, recorded as a function of delay time between photolysis and probe lasers at 50 K are displayed in Figure 1.

These curves were readily described by a functional fit of the form

$$I_D = A\{exp(-k_{L(D)}t) - exp(-k't)\} \qquad (1)$$

where $k' = k[D_2] + k_{L(C)}$, where $k$ represents the second order rate constant for the C($^1$D) + D$_2$ reaction, $k_{L(C)}$ represents the first-order loss of C($^1$D) by other processes such as secondary reactions and diffusion, [D$_2$] is the D$_2$ concentration, $t$ is time and $A$ is the signal amplitude. The first term in expression (1) represents D-atom losses with a first-order rate constant $k_{L(D)}$.

Decay curves were recorded at several $D_2$ concentrations for any particular experiment, to yield a wide range of values for the first-order rate constant $k'$. These $k'$ values were then plotted against the corresponding $D_2$ concentration to yield the second-order rate constant at a given temperature from a weighted linear least squares fit to the data (weighting was performed using the statistical uncertainties generated by the biexponential fitting procedure outlined above). Examples of such plots obtained at 50 K and at 296 K are shown in Figure 2.

The measured rate constants for the $C(^1D) + D_2$ reaction are summarized in Table 1 and displayed as a function of temperature in Figure 3 alongside the present RPMD rate constants and the previous results, both theory and experiment for this process. We also include our previous experimental results for the $C(^1D) + H_2$ reaction[43] for comparison. Error bars on the present experimental values were derived by combining the statistical uncertainties obtained from second-order fits such as those shown in Figure 2 with an estimated systematic error of 10 % of the nominal rate constant value. This systematic error was considered to potentially arise from calibration errors in various pieces of equipment such as mass-flow controllers and pressure gauges in addition to possible errors in the flow density and velocity calculations.

While the measured rate constants for the $C(^1D) + D_2$ reaction display only a very slight negative temperature dependence (independent of temperature considering the experimental error bars), it is interesting to note that our earlier measurements of the $C(^1D) + H_2$ reaction showed a slight positive temperature dependence [43]. Although these effects are relatively minor, the differences could be due to the use of non-equilibrated mixtures of ortho/para-$H_2$,-$D_2$ in both studies. If the reactivity of the ortho and para forms of these isotopologues towards $C(^1D)$ is significantly different, this should induce the largest impact on the measured rate at the lowest temperatures where deviations from equilibrium are greatest. In the case of the $C(^1D) + H_2$ reaction, for such an effect to occur would require the $j = 0$ level of $H_2$ (para-$H_2$) to react more rapidly than the $j = 1$ level (ortho-$H_2$) at the lowest temperatures. Consequently, the measured rates would be lower than expected due to an excess of ortho-$H_2$ in these earlier experiments. Although an earlier time-dependent wavepacket study of the C($^1$D) + $H_2$ reaction over uncoupled $^1A'$ and $^1A''$ PESs [22] concluded that the rotational state specific rate constant for the $j = 0$ level was approximately 15 % higher than the corresponding $j = 1$ level at 300 K, the rate constant for the $j = 0$ level is seen to increase at 100 K whereas the value for the $j = 1$ level remains constant. As a result, it may be possible that the relative reactivity trends are inverted at even lower temperatures. In the case of the C($^1$D) + $D_2$ reaction, a time-dependent wavepacket study indicated that the $j = 0$ (ortho) and $j = 1$ (para) levels of $D_2$ displayed the same reactivity at 300 K [32]. Unfortunately, no further information was provided with regard to the temperature dependence of the state specific rates for this system, making it impossible to determine whether the present measurements deviate from the expected values at equilibrium. Nevertheless, the temperature dependence of the RPMD results (which assume Boltzmann statistics for the rotational levels of $D_2$) is in good agreement with the experimentally determined one, suggesting that the relative reactivity of $j = 0$ and $j = 1$ levels of $D_2$ could be similar below 100 K. We will continue the discussion of the results presented in Figure 3 in Section 4.

**3 Theoretical Results**

The present theoretical simulations of the title reaction were performed using the RPMD method. This recently proposed[45] semiclassical method takes into account quantum mechanical effects of nuclear motions via the so-called "classical isomorphism" between the quantum mechanical formalism of statistical mechanics and the classical one for the ring polymer with beads being classical copies of the original system connected with neighbors in the necklace via fictitious temperature-dependent harmonic interaction. The classical isomorphism is exact for various static properties but it was also shown to provide a reliable approximation to real-time Kubo-transformed correlation functions responsible for various dynamical properties[50]. It has been found that RPMD provides very reliable estimates for real-time correlation functions used to calculate thermal rate constants[46,51,52]. Using simple gas-phase chemical reactions[43,53-69], the method was thoroughly compared with rigorous quantum mechanical results (when possible), other dynamics approximations and experiment and its accuracy and reliability has been confirmed as described in a recent review by one of us (YVS). In particular, RPMD exhibits very reliable and predictable behavior for barrierless chemical reactions with deviation from the rigorous quantum dynamics results within the convergence error of the computational procedure (typically below 15 %) [59,60]. As mentioned in the introduction, RPMD demonstrated an excellent agreement with experiment in our previous calculations for the C($^1$D) + H$_2$ reaction in the temperature range 50-300 K [43].

In the present RPMD calculations we used the same computational strategy as in our previous study of the C($^1$D) + H$_2$ reaction. All calculations were performed using the RPMDrate code [70]. The parameters of the simulations are summarized in Table 2. They are similar to the previous parameters for the C($^1$D) + H$_2$ system, except that fewer beads were required to converge the RPMD rate constants for the title D-transfer reaction. As previously [43], we used the multi-reference configuration-interaction PESs of Bañares *et al.* [17] for the ground state ($^1A'$) and of Bussery-Honvault *et al.* [21] for the first excited state ($^1A''$). Both PESs were obtained from *ab initio* data and are of insertion type with deep wells (4.32 and 3.46 eV relative to the reactants for $^1A'$ and $^1A''$, respectively) and barrierless minimum energy paths in the entrance channel (perpendicularly constrained approach and bent approach of C towards D$_2$ around 60° for $^1A'$ and

$^1A''$, respectively). As for the C($^1$D) + H$_2$ reaction, we used the $^1A'$ and $^1A''$ saddle points to initiate the RPMDrate calculations, though we note that the RPMD rate is rigorously independent of our choice of the initial transition state structure [52].

The results of the RPMDrate simulations are summarized in Figure 4 and 5 and in Table 3. Figure 4 shows that for both PESs, ring polymer free energy profiles demonstrate small thermodynamically induced barriers near the entrance to the well which diminish as we decrease the temperature. The $^1A''$ surface exhibits a more pronounced decrease. Figure 5 shows that for the $^1A'$ PES, recrossing dynamics are significantly enhanced as the temperature is decreased. Though the free energy barrier decreases with the temperature, this does not compensate the decrease due to recrossing dynamics. As a result, the RPMD rate coefficient for the $^1A'$ PES decreases with decreasing temperature, see Table 3. For the $^1A''$ PES, the temperature dependences of the ring polymer free energy barriers and the transmission coefficient are qualitatively the same, though the decrease in the plateau values of the ring polymer transmission coefficients is less pronounced and, as a result, the RPMD rate constant increases with the temperature. As a result, the rate constants for the $^1A'$ PES are higher than those for the $^1A''$ one only at 300 K. At temperatures below 75 K, the rate constants for the $^1A''$ PES are twice as large as those for the $^1A'$ PES.

When compared to the previous theoretical results for the C($^1$D) + H$_2$ chemical reaction [43], we note that the above temperature tendencies are similar to the previously observed ones. The final RPMD rate constants are smaller for C($^1$D) + D$_2$ and, as expected this difference decreases with decreasing temperature (see also Figure 3). However, the kinetic isotope effect depends on the PES: the rate constants are smaller for the C($^1$D) + D$_2$ reaction compared to the C($^1$D) + H$_2$ reaction by 20-40 % for the $^1A'$ PES and by only 13-16% for the $^1A''$ PES. The present results show that the $^1A''$ PES contribution is more important for the C($^1$D) + D$_2$ reaction, in line with the previous quantum mechanical time dependent wavepacket results [22,32]. Clearly, the chemical dynamics of the title reaction exhibit high sensitivity to the underlying PESs which therefore require very accurate treatment.

**4 Discussion and Conclusions**

Following our previous work on the C($^1$D) + H$_2$ chemical reaction [43], in this paper we performed a joint experimental and theoretical study of its deuterated counterpart, C($^1$D) + D$_2$, using the same experimental and theoretical techniques. Measurements were made using a continuous supersonic flow apparatus over the 50 – 296 K range. Due to the lack of an appropriate method for the direct detection of C($^1$D) atoms, rate constants were derived instead from product D($^2$S) atom formation curves. Theoretical results were obtained using novel rate theory based on ring polymer molecular dynamics (RPMD) which has deserved high confidence in the previous intensive examination of the method using prototype chemical reactions [43], including those with profiles of potential energy surface similar to the title reaction.

Figure 3 shows that the present RPMD and experimental results are in extremely good agreement, with the former being higher than the latter by only 15-25 % but correctly reproducing the very slight temperature dependence observed experimentally. Such deviations are slightly higher than the convergence error of the computational procedure [43,70] or the typical error of RPMD observed for prototype insertion chemical reactions [43,59,60]. We note that inaccuracy in the present RPMD results can be attributed to the sharp probability resonances found for the C($^1$D) + D$_2$ reaction [32] and which are not taken into account by the RPMD formalism [43]. It can also be attributed to inaccuracies of the underlying potential energy surfaces (PESs) and the adiabatic limit used in the present description of the electronic structure (i.e., the two PESs were treated as being uncoupled), as was discussed in our previous study of C($^1$D) + H$_2$ [43]. However, we note that the previous attempts to include non-adiabatic effects in calculations of the C($^1$D) + H$_2$ reaction led to a strong reduction in the overall rate constant and an inversion of the temperature dependence compared to the adiabatic limit [43]. Nonadiabatic RPMD studies for the C($^1$D) + X$_2$ reactions with X = H or D are clearly desirable in the future, when the nonadiabatic RPMD rate theory will become as robust and reliable as its original counterpart for single PESs.

Despite the small discrepancy, the present results demonstrate high consistency between theory and experiment. They confirm that both of the two lowest single PESs actively participate in the title reaction. In fact, the contribution from the first excited surface $^1A''$ prevails at temperatures below 300 K, being twice as large as the contribution from the $^1A''$ surface at 75-50 K. We note that the previous quantum mechanical time dependent wavepacket calculations, also obtained using two uncoupled PESs, [32] exhibit similar accuracy though underestimating experimental rates, as can be seen from Figure 3. Overall, the present RPMD results confirm the validity of this method for low-temperature studies of complex-forming reactions.

**Acknowledgments**

KMH acknowledges support from the French INSU/CNRS Programs 'Physique et Chimie du Milieu Interstellaire' (PCMI) and 'Programme National de Planétologie' (PNP). YVS thanks the European Regional Development Fund and the Republic of Cyprus for support through the Research Promotion Foundation (Project Cy-Tera ΝΕΑ ΥΠΟΔΟΜΗ/ΣΤΡΑΤΗ/0308/31).

**Table 1** Measured rate constants for the $C(^1D) + D_2$ reaction.

| T / K | $N^b$ | $[D_2]$ / $10^{14}$ molecule cm$^{-3}$ | $k_{C(1D)+D2}$ / cm$^3$ molecule$^{-1}$ s$^{-1}$ |
|---|---|---|---|
| 296 | 40 | 0.5 - 12.3 | $(1.9 \pm 0.2)^c \times 10^{-10}$ |
| 127 ± 2$^a$ | 40 | 0.4 - 9.9 | $(2.0 \pm 0.2) \times 10^{-10}$ |
| 75 ± 2 | 41 | 0.4 - 5.7 | $(2.0 \pm 0.2) \times 10^{-10}$ |
| 50 ± 1 | 35 | 0.6 - 7.3 | $(2.1 \pm 0.2) \times 10^{-10}$ |

$^a$Uncertainties on the calculated temperatures represent the statistical (1σ) errors obtained from Pitot tube measurements of the impact pressure; $^b$Number of individual measurements; $^c$Uncertainties on the measured rate constants represent the combined statistical and systematic errors as explained in the text.

**Table 2** Input parameters for the RPMDrate calculations of the C($^1$D) + D$_2$ reaction. The explanation of the format of the input file can be found in the RPMDrate code manual (see http://rpmdrate.cyi.ac.cy).

| Parameter | Potential Energy Surface | | Explanation |
|---|---|---|---|
| | $^1A'$ | $^1A''$ | |
| Command line parameters | | | |
| Temp | 300 | | Temperature (K) |
| | 127 | | |
| | 75 | | |
| | 50 | | |
| $N_{beads}$ | 128 (50 and 75 K); 64 (127 K); 48 (300 K) | | Number of Beads |
| Dividing surface parameters | | | |
| $R_\infty$ | 15 $a_0$ | 20 Å | Dividing surface parameter (distance) |
| $N_{bond}$ | 1 | 1 | Number of forming and breaking bonds |
| $N_{channel}$ | 2 | 2 | Number of equivalent product channels |
| X($^1$D) | (1.635 $a_0$, 1.302 $a_0$, 0.000 $a_0$) | (-2.450 $a_0$, 0.000 $a_0$, 0.000 $a_0$) | Cartesian coordinates (x, y, z) |
| H | (0.000 $a_0$, 0.000 $a_0$, 0.000 $a_0$) | (0.000 $a_0$, 0.000 $a_0$, 0.000 $a_0$) | of the intermediate geometry |
| H | (3.270 $a_0$, 0.000 $a_0$, 0.000 $a_0$) | (1.900 $a_0$, 0.000 $a_0$, 0.000 $a_0$) | |
| Thermostat | 'Andersen' | 'Andersen' | Thermostat option |
| Biased sampling parameters | | | |
| $N_{windows}$ | 111 | 111 | Number of Windows |
| $\xi 1$ | -0.05 | -0.05 | Center of the first window |
| $d\xi$ | 0.01 | 0.01 | Window spacing step |
| $\xi N$ | 1.05 | 1.05 | Center of the last window |
| $dt$ | 0.0001 | 0.0001 | Time step (ps) |
| $ki$ | 2.72 | 2.72 | Umbrella force constant ((T/K) eV) |
| $N_{trajectory}$ | 200 | 200 | Number of trajectories |
| $t_{equilibration}$ | 20 | 20 | Equilibration period (ps) |
| $t_{sampling}$ | 100 | 100 | Sampling period in each trajectory (ps) |
| $N_i$ | $2 \times 10^8$ | $2 \times 10^8$ | Total number of sampling points |
| Potential of mean force calculation | | | |
| $\xi_0$ | 0.00 | 0.00 | Start of umbrella integration |
| $\xi_\ddagger$ | 0.604 (300 K)[a] | 0.915 (300 K)[a] | End of umbrella integration |
| | 0.448 (127 K)[a] | 0.807 (128 K)[a] | |
| | 0.398 (75 K)[a] | 0.765 (77 K)[a] | |

|  | 0.361 (50 K)[a] | 0.754 (50 K)[a] |  |
| --- | --- | --- | --- |
| $N_{bins}$ | 5000 | 5000 | Number of bins |
| Recrossing factor calculation |  |  |  |
| $dt$ | 0.0001 | 0.0001 | Time step (ps) |
| $t_{equilibration}$ | 20 | 20 | Equilibration period (ps) in the constrained (parent) trajectory |
| $N_{totalchild}$ | 200000 | 200000 | Total number of unconstrained (child) trajectories |
| $t_{childsampling}$ | 3 | 3 | Sampling increment along the parent trajectory (ps) |
| $N_{child}$ | 100 | 100 | Number of child trajectories per one initially constrained configuration |
| $t_{child}$ | 2 | 4 | Length of child trajectories (ps) |

[a]Detected automatically by RPMDrate.

**Table 3** Summary of the RPMDrate calculations for the C($^1$D) + D$_2$ reaction over the $^1A'$ and $^1A''$ PESs at 50, 75, 127 and 300 K: $k_{QTST}$ - centroid-density quantum transition state theory rate coefficient; $\kappa_{plateau}$ - ring polymer transmission coefficient; $k_{RPMD}$ – RPMD rate coefficient. The final RPMD rate constants (last column) are corrected by the electronic partition function $Q_{el}$ = 1/5. The parentheses denote powers of ten.

| T(K) | $k_{QTST}$ | $\kappa_{plateau}$ | $k_{RPMD}$ | $k_{RPMD}$(corrected by $Q_{el}$) |
| --- | --- | --- | --- | --- |
| PES $^1A'$ | | | | |
| 300 | 1.20(-09) | 0.489 | 5.84(-10) | 1.17(-10) |
| 127 | 9.43(-10) | 0.529 | 4.99(-10) | 9.98(-11) |
| 75 | 7.48(-10) | 0.594 | 4.44(-10) | 8.88(-11) |
| 50 | 6.16(-10) | 0.652 | 4.02(-10) | 8.04(-11) |
| PES $^1A''$ | | | | |
| 300 | 2.08(-09) | 0.247 | 5.14(-10) | 1.03(-10) |
| 127 | 2.51(-09) | 0.277 | 6.95(-10) | 1.39(-10) |
| 75 | 2.51(-09) | 0.323 | 8.11(-10) | 1.62(-10) |
| 50 | 2.47(-09) | 0.358 | 8.85(-10) | 1.77(-10) |

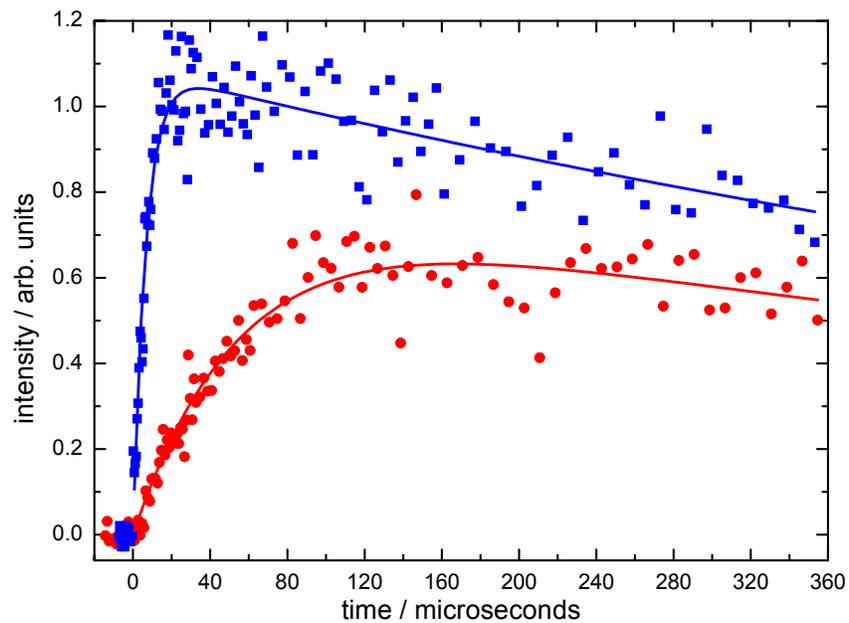

**Figure 1** Typical D($^2$S) atom formation curves from the C($^1$D) + D$_2$ → CD + D reaction recorded at 50 K. (Blue solid squares) [D$_2$] = 6.8 × 10$^{14}$ molecule cm$^{-3}$; (red solid circles) [D$_2$] = 6.0 × 10$^{13}$ molecule cm$^{-3}$. Solid lines represent biexponential fits to the data of the form given by equation (1).

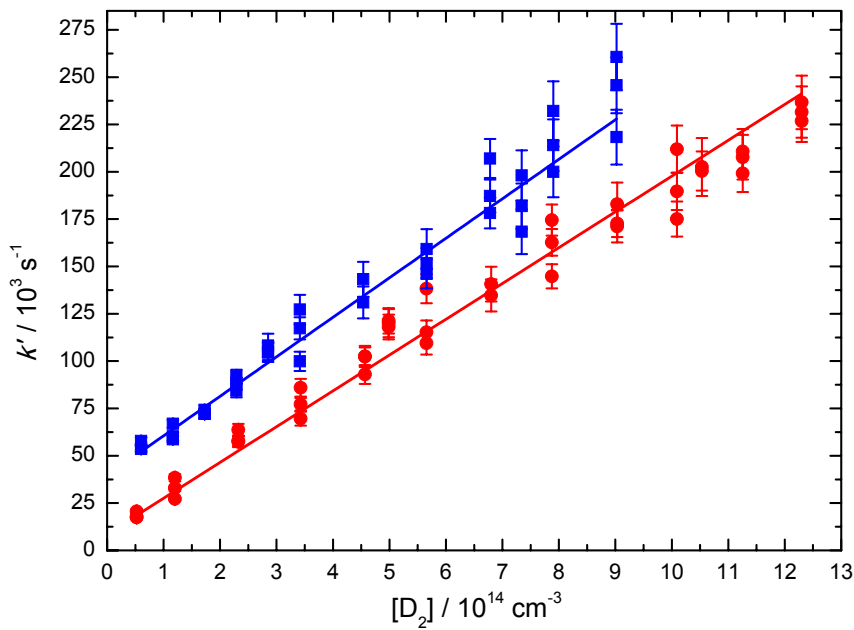

**Figure 2** Pseudo-first order rate constants for the C($^1$D) + D$_2$ reaction as a function of [D$_2$], recorded at 296 K (red solid circles) and at 50 K (blue solid squares). The data at 50 K are displaced upwards by 40000 s$^{-1}$ for clarity. Error bars represent the statistical uncertainties of the pseudo-first-order rate constants derived from the biexponential fitting procedure. Second-order rate constants were determined by weighted fits to the individual datasets (solid blue and red lines) as described in the text.

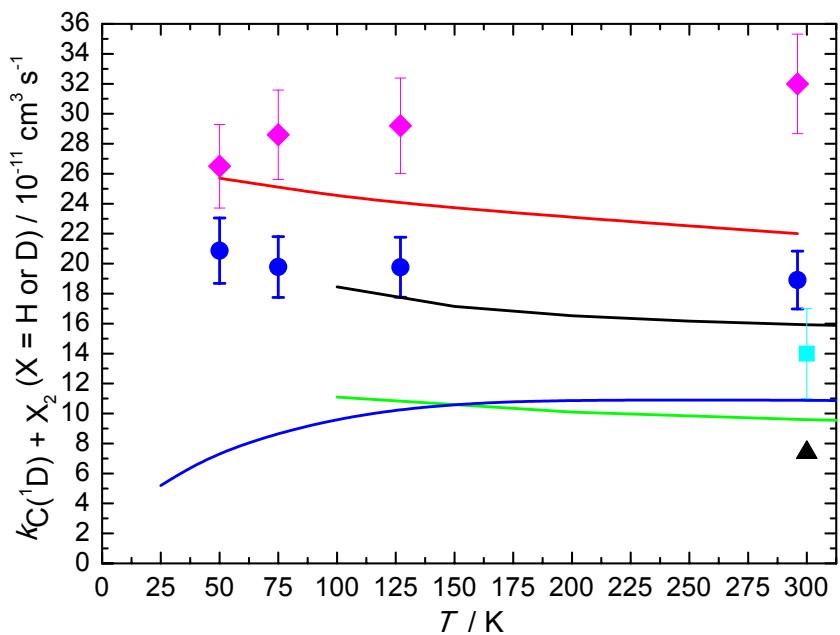

**Figure 3** Rate constants for the C($^1$D) + X$_2$ reactions as a function of temperature (X = H or D). Theoretical values for the C($^1$D) + D$_2$ reaction: (solid black line) Defazio *et al.* [32]; ((solid green line) Joseph *et al.* [18]; (solid blue line) Sun *et al.* [29]; (black triangle) Lin and Guo [24]; (solid red line) this work, RPMD method. Experimental values for the C($^1$D) + D$_2$ reaction: (light blue square) Sato *et al.*[42]; (dark blue circles) this work, supersonic flow reactor; (purple diamonds) rate constants for the C($^1$D) + H$_2$ reaction from Hickson *et al.* [43]. Uncertainties on the present experimental results are the combined statistical (1σ) and estimated systematic errors (10 %).

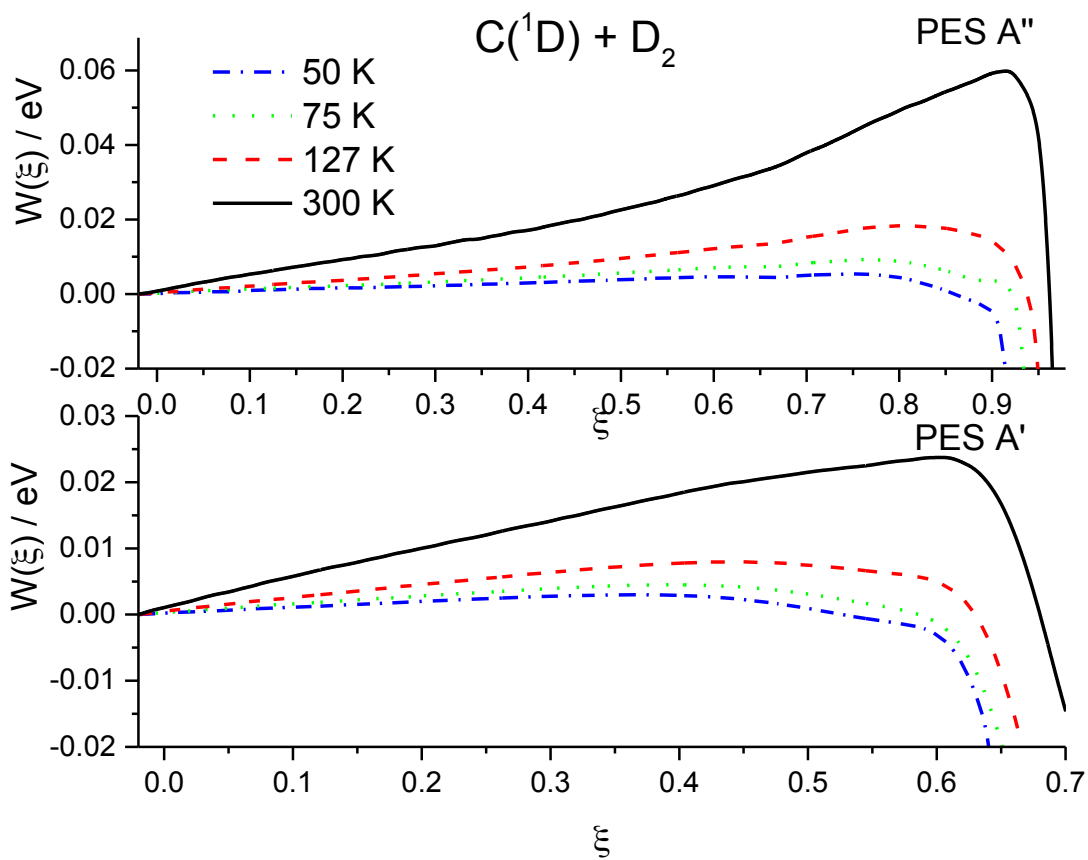

**Figure 4** Ring polymer potential of mean force (free energy) for the C($^1$D) + D$_2$ reaction at 50, 75, 127 and 300 K over two potential energy surfaces $^1$A′ (lower panel) and $^1$A″ (upper panel).

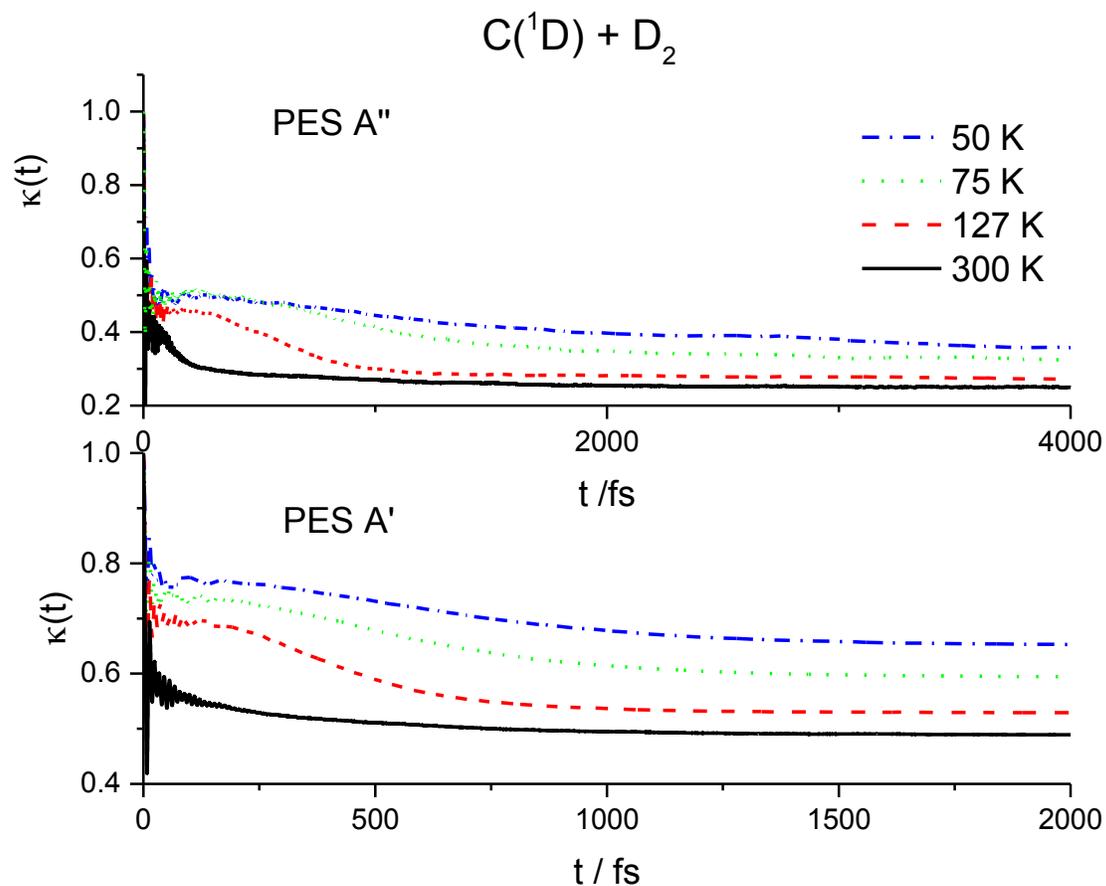

**Figure 5** Ring polymer transmission coefficients for the C($^1$D) + D$_2$ reaction at 50, 75, 127 and 300 K over two potential energy surfaces $^1$A′ (lower panel) and $^1$A″ (upper panel).

**TOC Entry**

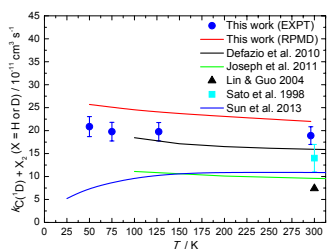